\documentclass[12pt]{article}
\usepackage{geometry} % see geometry.pdf on how to lay out the page. There's lots.
\usepackage{graphicx} % for graphics
\usepackage{amsmath} % for AMS style symbols like \mathbb
\usepackage{amsfonts} %
\usepackage{cite}          % for nice numbering of the citations
\usepackage{hyperref}  % for hyperlinks, esp. at the bibliography

\usepackage{tikz}
\usepackage{pgfplots}
\usetikzlibrary{decorations.markings}
\title{Conformal Quantum Mechanics and Sine-Square Deformation}
\author{Tsukasa TADA}

\date{} % uncomment this line to display a designated date
\begin{document}
\maketitle
\centerline{\it RIKEN Nishina Center for Accelerator-based Science \&} 
\centerline{\it Interdisciplinary Theoretical and Mathematical Sciences (iTHEMS)}
\centerline{\it Wako, Saitama 351-0198, Japan}
\begin{abstract}
We revisit conformal quantum mechanics (CQM)  from the perspective of sine-square deformation (SSD) and the entanglement Hamiltonian. The operators %that correspond 
related to SSD and the entanglement Hamiltonian are identified. Thus, the nature of SSD and the entanglement can be discussed in a much simpler CQM setting than higher-dimensional field theories.
\end{abstract}
%\tableofcontents

%\section{Introduction}
In \cite{Ishibashi:2015jba,Ishibashi:2016bey}, it was shown that sine-square deformation (SSD) \cite{SSD} for two-dimensional (2d) conformal field theory (CFT) \cite{Katsura:2011ss} can be understood by introducing a new quantization scheme called  ``dipolar quantization.''\footnote{See Refs. \cite{Proof1,Proof2} for earlier studies on SSD and Refs. \cite{Okunishi:2015dfa,Okunishi:2016zat,Tamura:2017vbx} for more recent studies. References \cite{Tada:2014kza,Tada:2014jps} study SSD in the context of string theory and conformal field theory.}
The basic idea was  generalized in Ref. \cite{Wen:2016inm} to incorporate the entanglement Hamiltonian and other interesting deformations of 2d CFT.
In this Letter, we examine whether the idea of dipolar quantization is applicable to the one-dimensional (1d) case, which is called conformal quantum mechanics (CQM). CQM was first studied in the seminal paper by de Alfaro, Fubini, and Furlan \cite{deAlfaro:1976vlx}.

To put the problem in perspective, let us consider a scalar field $\phi(x)$ on general d-dimensional flat spacetime $x^{\mu} (\mu=0, \dots, d-1)$ following the argument presented in Ref. \cite{deAlfaro:1976vlx}. Suppose $\phi(x)$ transforms under the scale transformation $x_{\mu} \rightarrow x'_{\mu}=\lambda x_{\mu}$ as
\begin{equation}
%\phi(\lambda x^{\mu})=\lambda^{-\frac{d-2}{2}}\phi(x^{\mu}).
\phi(x^{\mu})\rightarrow\phi'(x'^{\mu})=\phi'(\lambda x^{\mu})=\lambda^{-\frac{2d}{2}}\phi(x^{\mu})
\end{equation}
A simple invariant action for $\phi(x)$  can be obtained as
\begin{equation}
S=\int \prod_{\mu}dx^{\mu} \frac12\left(\partial_{\nu}\phi\partial^{\nu}\phi-g\phi^{\frac{2d}{d-2}}\right) , \label{eqn:S4d}
\end{equation}
where $g$ is the dimensionless coupling constant.
%If we consider a scalar field $\phi(x)$ on d-dimensional flat spacetime $x^{\mu} (\mu=0, \dots, d-1)$  that $\phi$ transforms under the scale transformation as
%\begin{equation}
%\phi(x^{\mu})\rightarrow\phi'(x'^{\mu})=\phi'(\lambda x^{\mu})=\lambda^{-\frac{2d}{2}}\phi(x^{\mu})
%\end{equation}
Because scale invariance implies  conformal invariance in most cases \cite{Nakayama:2013is}, this action provides a good starting point. In addition, Eq. (\ref{eqn:S4d})  suggests the difficulty in the Lagrangian formalism for $d=2$ case, in which the energy momentum tensor is taken as the basis of the theory rather than the lagrangian.

The case of interest here  is $d=1$, which has the following Lagrangian:
\begin{equation}
L=\frac12 (\dot{q}(t))^{2}-\frac{g}{2}\frac{1}{q(t)^{2}}, \label{eqn:CQML}
\end{equation}
where $t$ is the 1d ``spacetime'' coordinate. We also changed the notation of the ``field'' from $\phi$ to $q(t)$ because we are now dealing with a quantum mechanical system. We can then show that the Lagrangian (\ref{eqn:CQML}) possesses the following symmetry:
\begin{eqnarray}
t &\rightarrow & t'=\frac{at+b}{ct+d}, \ \ ad-bc=1, \label{eqn:tmobius}\\
q(t) &\rightarrow&q'(t')=\frac{1}{ct+d}q(t),
\end{eqnarray}
which is a larger symmetry than scale invariance and  translational invariance combined. In fact, it is 1d conformal symmetry, as we anticipated.

The transformation (\ref{eqn:tmobius}) for $t$ can be conveniently decomposed into the following three components:\\
\noindent{\it Translation} $a=d=1$ and $c=0$ lead to
\begin{equation}
t  \rightarrow  t+b.   \label{eqn:translation}
\end{equation}
{\it Dilatation} $a=1/d$ and $b=c=0$ lead to
\begin{equation}
t  \rightarrow  a^{2}t.   \label{eqn:dilatation}
\end{equation}
{\it Special Conformal Transformation(SCT)} $a=d=1$ and $b=0$ lead to
\begin{equation}
t \rightarrow  \frac{t}{ct+1}.\label{eqn:sct}
\end{equation}
The infinitesimal version of transformations (\ref{eqn:translation}) - (\ref{eqn:sct}) of the above three  can be represented in terms of the differential operators as follows.
\begin{eqnarray}
{(Time) Translation} \ \ \frac{d}{dt} &\equiv &  P_{0}, \\
Dilatation \ \  t\frac{d}{dt} & \equiv &D, \\
SCT \ \ t^{2}\frac{d}{dt} &\equiv &K_{0}.
\end{eqnarray}
These operators form a closed algebra,
\begin{equation}
[D, K_{0}]=K_{0}, \ [D,P_{0}]=-P_{0}, \ [P_{0}, K_{0}]=2D, \label{eqn:commDKP}
\end{equation}
which is readily isomorphic to  $sl(2, \mathbb{R})$ algebra or, equivalently, the subalgebra formed by  the three Virasoro generators $L_{1}, L_{0},$ and $L_{-1}$:
\begin{equation}
[L_{0}, L_{-1}]=L_{-1}, \ [L_{0}, L_{1}]=-L_{1}, \ [L_{1}, L_{-1}]=2L_{0}. \label{eqn:Virsub}
\end{equation}

The time-translation generator $P_{0}$ should be identified with the Hamiltonian
\begin{equation}
H= \frac12 p(t)^{2}+\frac{g}{2}\frac{1}{q(t)^{2}}, \label{eqn:CQMH}
\end{equation}
where $p$ is the canonical momentum
%%the following ADDED on Revision
if one regards the commutation relation in (\ref{eqn:commDKP}) as the Poisson bracket \footnote{Note the difference in the factor $i$ of the algebra (\ref{eqn:commDKP}), 
compared to that in Ref. \cite{deAlfaro:1976vlx}.}
.
Equation (\ref{eqn:CQMH}) was directly derived from Lagrangian (\ref{eqn:CQML}). Using the symplectic structure, the rest of the generators may be expressed in terms of $q$ and $p$:
\begin{equation}
K_{0}=\frac12 q(t)^{2}, \label{eqn:CQMK}
\end{equation}
\begin{equation}
D=-\frac14\left(p(t)q(t)+q(t)p(t) \right). \label{eqn:CQMD}
\end{equation}
In Eq. (\ref{eqn:CQMD}) we employed symmetrization in anticipation of quantization.

In Ref. \cite{deAlfaro:1976vlx}, de Alfaro, Fubini, and Furlan introduced the new  operator
\begin{equation}
R \equiv \frac12 \left(aP_{0}+\frac{1}{a}K_{0}\right),
\end{equation}
where $a$ is a constant with the dimensions of time, along with two other operators. %Without loss of generality, we set $a$ to be unity hereafter.
Then, $R$ was proposed to supersede $H$ as the time-translation operator, or the Hamiltonian.

The distinction between the operator $R$ and the original Hamiltonian $H=P_{0}$ is best clarified from the symmetry viewpoint \cite{deAlfaro:1976vlx,Ishibashi:2016bey}. First, the (quadratic) Casimir invariant for  $sl(2, \mathbb{R})$ algebra is
%\begin{equation}
%C_{(2)}=\left( L_{0}\right)^{2}-\frac12L_{-1}L_{1}-\frac12L_{1}L_{-1}=D^{2}-\frac12K_{0}P_{0}-\frac12P_{0}K_{0}. \label{eqn:CasOp}
%\end{equation}  %%Note the reverse sign
\begin{equation}
C_{(2)}=\frac12L_{-1}L_{1}+\frac12L_{1}L_{-1}-\left( L_{0}\right)^{2}=\frac12K_{0}P_{0}+\frac12P_{0}K_{0} -D^{2}. \label{eqn:CasOp}
\end{equation}
Therefore, for any adjoint action of  $sl(2, \mathbb{R})$ algebra on the linear combination of the generators,
\begin{equation}
x^{(0)}L_{0}+x^{(1)}L_{1}+x^{(-1)}L_{-1}   \longrightarrow x'^{(0)} L_{0}+x'^{(1)}L_{1}+x'^{(-1)}L_{-1} ,
\end{equation}
 the following combination remains unchanged \footnote{%It may appear that we have simply translated the numerical coefficients of the quadratic form in Eq. (\ref{eqn:CasOp} ) to those in Eq. (\ref{eqn:CasCoeff}). Note, however, that the coefficients in these equations should be the components of inverse matrices to each other. The exact match in the numerical coefficients in Eq. (\ref{eqn:CasOp} ) and Eq. (\ref{eqn:CasCoeff}) is merely coincidence due to the special structure of the quadratic forms in question.}
Note that  the numerical coefficients of the quadratic form in Eqs. (\ref{eqn:CasOp}) and (\ref{eqn:CasCoeff}) are components of matrices that are inverse of each other.}:
\begin{equation}
2x^{(1)}x^{(-1)}+2x^{(-1)}x^{(1)} -\left( x^{(0)}\right)^{2}=4x'^{(1)}x'^{(-1)}-\left( x'^{(0)}\right)^{2} \equiv c^{(2)}\label{eqn:CasCoeff}
\end{equation}
In terms of the coefficients $x^{(0)}, x^{(1)},$ and $x^{(-1)}$, the operator $R$ is expressed as
\begin{equation}
R: x^{(0)}=0, x^{(1)}=\frac{a}{2},  x^{(-1)}=\frac{1}{2a},
\end{equation}
and the expression for the original Hamiltonian $H$ (or $P_{0}$) is
\begin{equation}
H: x^{(0)}=0, x^{(1)}=1,  x^{(-1)}=0.
\end{equation}
Putting these coefficients into  $c^{(2)}$ defined in Eq. (\ref{eqn:CasCoeff}),  we immediately find
\begin{equation}
c^{(2)}=1 \quad \mbox{for $R$}, \label{eqn:c2R}
\end{equation}
and 
\begin{equation}
c^{(2)}=0 \quad \mbox{for $H$}.
\end{equation}
These results imply that one cannot connect $R$ and $H$ by any adjoint action of $sl(2, \mathbb{R})$, nor by its exponentiation, $SL(2, \mathbb R)$. In this sense, operators $R$ and $H$ are disconnected.

Now, note the absence of  constant $a$ in expression (\ref{eqn:c2R}), which infers that $a$ can be changed numerically by an adjoint action of $sl(2, \mathbb{R})$ or $SL(2, \mathbb R)$ action on the operator $R$. In  fact, an infinitesimal change in $a\rightarrow a(1-\epsilon)$ evokes the commutation with $D$ as
\begin{eqnarray}
R\xrightarrow{a\rightarrow a(1-\epsilon)}  \frac12 \left(a(1-\epsilon)P_{0}+\frac{1}{a(1-\epsilon)}K_{0}\right)&=&R+\frac12\left(-aP_{0}+\frac{K_{0}}{a}\right)\epsilon \nonumber 
\\&=&R+[D,R]\epsilon.
\end{eqnarray}
Thus, different values of $a$ in $R$ are connected by the action of $D$. Two other actions  can be applied to
$R$, namely, $P_{0}$ and $K_{0}$, which would produce terms corresponding to $D$ and yield a nonzero $x^{(0)}$ coefficient. Hereinafter, we assume $a$ to be unity for the sake of simplicity.

We then ask if any class of operators is connected to $H$ by the action of $SL(2, \mathbb{R})$. Apparently, the answer is affirmative because the following operator $H^{(a,b)}$
\begin{equation}
H^{(a,b)}: x^{(0)}=\pm2\sqrt{ab}, x^{(1)}=a,  x^{(-1)}=b,  \ \ \  \mbox{for} \  \ ab \geq 0,   \label{eqn:Hab}
\end{equation}
yields $c^{(2)}=0$ as does $H$, which can be written in the above notation as
\begin{equation}
H=H^{(1,0)}.
\end{equation}
$H^{(a,b)}$ is explicitly written as
\begin{equation}
H^{(a,b)}=aP_{0}+bK_{0}\pm2\sqrt{ab}D,
\end{equation}
%\begin{equation}
%H^{(a,b)}=aP_{0}+bK_{0}-2\sqrt{ab}D.
%\end{equation}
or, in terms of the canonical variables,
\begin{equation}
H^{(a,b)}=\frac{a}{2} p(t)^{2}+\frac{ag}{2}\frac{1}{q(t)^{2}}+\frac{b}{2} q(t)^{2} -\sqrt{ab}p(t)q(t) ,
\end{equation}
where, without loss of generality, we have chosen one of the double signs that appeared in Eq. (\ref{eqn:Hab}).

The transformation between $H^{(1,0)}$ and $H^{(a,b)}$ can be interpreted in terms of classical mechanics because we have designed the system so that it accommodates  conformal symmetry. In fact, the transformation can be achieved by changing the canonical coordinates as follows:
\begin{equation}
\left\{\begin{array}{c}q(t) \rightarrow \frac{1}{\sqrt{a}} Q(t) \\p(t)\rightarrow \sqrt{a}P(t)-\sqrt{b}Q(t) \end{array}\right. ,
\end{equation}
where $P(t)$ and $Q(t)$ are the new canonical coordinates.
The generating function of the above canonical transformation is
\begin{equation}
W=\sqrt{a}q(t)P(t)-\frac{\sqrt{ab}}{2}q^{2}(t).
\end{equation}

%It would be instructive to recast the above discussion in the  $SL(2,R)$ perspective. $SL(2,R)$ is most naturally represented as the automorphism of the (upper) half plane with naturally hyperbolic metric. The introduction the hyperbolic metric is an option, of course but it should be as natural as we consider sphere with the positive metric. Then the automorphism is

Another class of generators yields negative $c^{(2)}$, the simplest of which is
\begin{equation}
{\bar R} \equiv H-K_{0} = \frac12 p(t)^2 + \frac12 \frac{g}{q(t)^2} - \frac12 q(t)^2.
\end{equation}
For ${\bar R}$, the coefficients are
\begin{equation}
{\bar R}: x^{(0)}=0, x^{(1)}=1,  x^{(-1)}=-1,
\end{equation}
which yield $c^{(2)}=-1$.
\footnote{${\bar R}$ corresponds to $-S$ in the notation of Ref. \cite{deAlfaro:1976vlx}.}

\begin{figure}[tbh]
\centering
\begin{tikzpicture}[domain=0.25:4,samples=60,>=latex%stealth
]
%  \draw[very thin,color=gray] (-0.1,-1.1) grid (3.9,3.9);
  \draw[->] (-0.2,0) -- (4.5,0) node[right] {$q$};
  \draw[->] (0,-1.2) -- (0,4) node[left] {$V(q)$};
  \draw[color=red,thick]    plot (\x,1/\x*1/\x/4+\x*\x/8)             node[right] {$c^{(2)}=1$};
  \draw[color=blue,thick]   plot (\x,1/\x*1/\x/4)    node[above] {$c^{(2)}=0$};
  \draw[color=teal,thick] plot (\x,1/\x*1/\x/4-\x*\x/8) node[right] {$c^{(2)}=-1$};
\end{tikzpicture}
\caption{Potential $V(q)$  for $c^{(2)}=1,0,$ and $-1$.
} \label{fig:potential}
\end{figure}
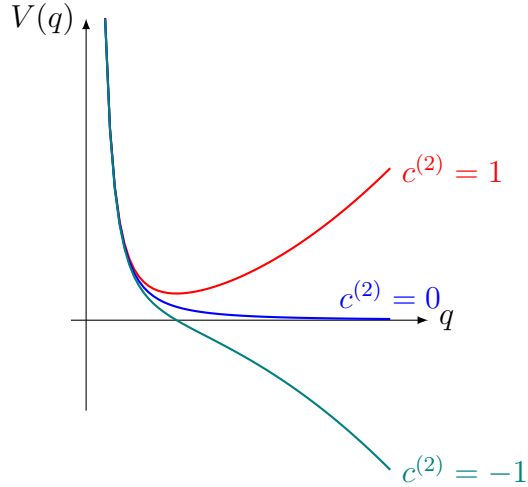

Now, each distinct class of $c^{(2)}$ can be conveniently represented by the following combination of coefficients:
\begin{equation}
x^{(0)}=0, x^{(1)}=1,  x^{(-1)}=\frac{c^{(2)}}{4}.
\end{equation}
The corresponding generator is
\begin{equation}
H+\frac{c^{(2)}}{4}K_{0}=\frac12 p^{2}+\frac{g}{2}\frac{1}{q^{2}}+\frac{c^{(2)}}{8} q^{2}.
\end{equation}
Because the above generator resembles the ordinary Hamiltonian, it is clarifying to draw the graph of the potential $V(q)=\frac{g}{2}\frac{1}{q^{2}}+\frac{c^{(2)}}{8} q^{2}$ for each case
%%ADDED on REVISION
, where we presume $g > 0$
\footnote{For negative $g$, despite the apparent unbounded potential, the corresponding Schr\"odinger equation for $c^{(2)}=0$ has a stable solution up to $g=-\frac14$, similar to the Breitenlohner-Freedman bound \cite{Breitenlohner:1982bm} in higher dimensional AdS space. }
. 
Figure \ref{fig:potential} shows the potential for the cases where $c^{(2)}$ equals $1$, $0$, and $-1$, respectively. In the following, we investigate each case.

Reference \cite{deAlfaro:1976vlx} observed that the invariance of the Casimir invariant (\ref{eqn:CasOp}) is apparent from the expressions (\ref{eqn:CQMH}) - (\ref{eqn:CQMD}), if one imposes the commutation relation over $q$ and $p$ as $[q,p]=i\mathbb{I}$:
\begin{equation}
\frac12HK_{0}+\frac12K_{0}H-D^{2}=\left(\frac{g}{4}-\frac{3}{16}\right)\mathbb{I},
\end{equation}
where $\mathbb{I}$ is the identity operator of the (enveloping) algebra in question
\footnote{As noted in Ref. \cite{deAlfaro:1976vlx}, the case $g=0$ yields  particularly simple representations by the creation and annihilation operators $[a,a^{\dagger}]=1$, which are called  singleton representations. Despite the ostensible lack of enough structure to accommodate the symmetry, this is an example of a spectrum generating algebra, and the symmetry algebra is represented by the transitions between the different energy states \cite{Joseph:1974hr,Ramond:2010zz}.}.
Without fear of confusion, we also denote the parameter $\left(\frac{g}{4}-\frac{3}{16}\right)$ as $C_{(2)}$. Using the notation $C_{{(2)}}$, one obtains
% \begin{eqnarray}
%L_{1}L_{-1}&=&L^{2}_{0}+L_{0} - C_{(2)}\mathbb{I}, \\
%L_{-1}L_{1}&=&L^{2}_{0}-L_{0} - C_{(2)}\mathbb{I},
%\end{eqnarray}
\begin{equation}
L_{\pm1}L_{\mp1}=L^{2}_{0}\pm L_{0} - C_{(2)}\mathbb{I}  \label{eqn:L0square}
\end{equation}
which turns out to be useful for finding the eigenvalues of $L_{0}$.

Suppose  a normalized eigenstate vector $|E\rangle$ exists such that
\begin{equation}
L_{0}|E\rangle=E|E\rangle ,  \quad \langle E|E\rangle =1.
\end{equation}
It is then straightforward to show that one can construct eigenstates with eigenvalues $E\pm1$ by multiplying by $L_{\mp 1}$ because
 \begin{equation}
L_{0}\left(L_{\mp1}|E\rangle \right)=\left(L_{\mp1}L_{0}\right)|E\rangle+L_{\mp1}|E\rangle =(E\pm1)L_{\mp1}|E\rangle.
\end{equation}
We would like to normalize the  eigenstates obtained above,
\begin{equation}
L_{\mp1}|E\rangle\equiv c^{\pm}(E)|E\pm1\rangle,
\end{equation}
so that
\begin{equation}
\langle E\pm1|E\pm1\rangle=1.
\end{equation}
The normalization factor $c^{\pm}$ can be calculated using Eq. (\ref{eqn:L0square}), which yields
\begin{equation}
|c^{\pm}(E)|^{2}=\langle E|L_{\pm1}L_{\mp1}|E\rangle=\langle E|L^{2}_{0}\pm L_{0} - C_{(2)}\mathbb{I}|E\rangle =E^{2}\pm E-C_{(2)} \geq 0.
\end{equation}
This condition of positivity can be clearer if we introduce a common notation for the Casimir invariant $C_{(2)}=j(j-1)$ (we assume $j \geq 0$):
\begin{equation}
|c^{\pm}(E)|^{2}=E(E\pm 1)-j(j-1) \geq 0.
\end{equation}
We thus conclude that $E \geq j$ or $E \leq -j$, and from physical considerations, we prefer positive $E$. Finally, as the eigenvalues of $L_{0}$, we obtain
\begin{equation}
E=n+j,   \label{eqn:Reigenv}
\end{equation}
where $ n=0,1,2,3, \dots$\footnote{One might consider an extension of $sl(2, \mathbb{R})$ to the full Virasoro algebra on these eigenstates. See Ref. \cite{Kumar:1999fx} for a related discussion.}.
Because  $R$ can be identified with $L_{0}$, we obtain the system with a discrete spectrum  using $R$ as the Hamiltonian in stead of the original $H$
%footnote added to v.2 1/09/18
\footnote{With the identification of $R$ as $L_{0}$, $L_{\pm 1} = (H-K_{0})/2\pm i D$ \cite{deAlfaro:1976vlx}.}
. This is fairly evident from Fig. \ref{fig:potential} because $R$ corresponds to the case $c^{(2)}=1$, where the range of motion is apparently limited to a finite region.

Conversely, the case $c^{(2)}=0$ does  not exhibit discrete energy states because the motion of the particle is not confined by the potential. Instead, it has a continuous spectrum as discussed in detail in Ref. \cite{deAlfaro:1976vlx}. This emergence of the continuous spectrum compelled the authors of Ref. \cite{deAlfaro:1976vlx} to propose $R$ as the Hamiltonian of CQM instead of the original $H$, which corresponds to the case $c^{(2)}=0$.

However, we prefer to propose another interpretation of $H$ here: In light of SSD, we do not have to reject an operator just because it leads a continuous spectrum. In fact, this is the signature of SSD systems. Therefore, we propose to regard $H$ as the SSD Hamiltonian \footnote{See Ref. \cite{Okazaki:2017lpn} for a recent discussion on quantization using $H$.}. If we accept this interpretation, the relation between  radial quantization \cite{Fubini:1972mf}  and SSD in 2d conformal field theories \cite{Ishibashi:2015jba,Ishibashi:2016bey} naturally parallels that between $R$ and $H$. This interpretation offers a nice intuition on somewhat mysterious nature of the continuous spectrum of SSD: it stems from the runaway potential in the CQM case.

Next, we turn our attention to the case $c^{(2)}=-1$. Since the potential for this case is unbounded below, the system is unstable, no meaningful physical interpretation is apparent. However, the $sl(2,\mathbb{R})$ symmetry of the system enables the following analysis.

First, the generators $H, K_{0}, D,$ and $R$ allows another non-trivial identification with the Virasoro subalgebra:
\begin{eqnarray}
L'_0&=&\frac{1}{2i} \left( H-K_0\right)=\frac{1}{2i}{\bar R}, \\
L'_{-1}&=&-\frac12\left(H+K_0\right)-D=-\frac12 R -D,\\
L'_{1}&=&\frac12\left( H+K_0\right)-D=\frac12 R -D. 
\end{eqnarray}
The set of operators above satisfies the same commutation relations given in Eq. (\ref {eqn:Virsub}). Since the algebraic structure is the same, the eigenvalues for the operator $L'_0$ should be the same. However, the ``Hamiltonian'' in question is $\bar R$, not $L'_0$. The difference between $R$ and $L_{0}$ is the multiplication of the the imaginary unit $i$. Thus we find that the spectrum of  ${\bar R}$ is $2i$ times that of $R$.

What can we make of a ``Hamiltonian'' with pure imaginary eigenvalues?
Although imaginary eigenvalues appear unphysical, all these eigenvalues take the form $2iE_n$, where $E_n$ represent the eigenvalues of the ``physical'' Hamiltonian $R$, as explicitly shown in Eq. (\ref{eqn:Reigenv}).
If we rewrite $t\rightarrow \beta/2$, the ``time'' translation operator can be expressed as
\begin{equation}
\exp(it{\bar R})=\sum | n \rangle e^{-\beta E_n}  \langle n | ,
\end{equation}
and clearly corresponds to the thermal density matrix operator
\begin{equation}
\rho\equiv \frac{\exp({-\beta { R}}) }{ \text{Tr}\left[\exp({-\beta {R}})\right]}
\end{equation}
for the original system that quantized with $R$; the time translation evoked by $\bar R$ yields the thermal density matrix operator.

The relation between the density matrix operator $\rho$ and ${\bar R}$,
\begin{equation}
{\bar R} \sim -\frac{1}{i2t} \ln \rho %beta->2t corrected 04/12/18
\end{equation}
is that of the so-called modular Hamiltonian (for example, see \cite{Haag:1992hx}) except the extra $i$ factor. %
%FACTOR i difference
%Refer to Wen, Ryu, Ludwig and explain Minkowski vs Eulidean
%
Conventionally, the modular Hamiltonian, which is an hermitian operator, is used to construct a unitary operator with additional $i$ factor, but here the $i$ factor is already included. Therefore, simply exponentiating the $\bar R$ yields an unitary operator. Since the modular Hamiltonian is also called the entanglement Hamiltonian 
\cite{Hislop:1981uh,Haag:1992hx,Casini:2011kv,Blanco:2013joa,Li:2008kda,Peschel:2004iop,Cho:2017lvx,Cardy:2016fqc},
we  infer that this case $\bar R$ corresponds to the entanglement Hamiltonian. 
As a matter of fact, Wen, Ryu and Ludwig \cite{Wen:2016inm} pointed out in the study of 2d CFT that the  deformation of the Hamiltonian, which corresponds to $\bar R$ from the point of view of symmetry, yields the Rindler Hamiltonian, which is the extreme case of the entanglement Hamiltonian. %%%%%%%%%%%%%
Since they studied Euclidean field theory, the additional $i$-factor was absent for their case. %%%%%%%%%%%%%
%For the case at hand, the system has only one degree of freedom and there is no other degree of freedom to integrate out, thus, we naturally obtain the expression for the entire density matrix from the corresponding entanglement Hamiltonian for $d=1$. %Probably misleading argument

%CORRESPONDING CANONICAL TRANS ?
%
%Parabolic -> K, H(=P)
%Compact ->R
%Noncompact -> D, S
%
%vector field representation?
%
%ENTANGLEMENT HAMILTONIAN DIRECTION? ON the HALF PLANE?
%->C(2)=1  D
%significance with entanglement Hamiltonian?
%D+H -> C_(2)=1
%H 1/q^{2} so rather D+K     (q-p)^{2}-p^{2}     imaginary eigenvalue -> relation to KMS condition?  -> t compact ?
%pq linear term (p-q)^2 -q^2    p-q ->P  q->Q  runaway potential   significance?   imaginary eigenvalues possible
%Q->iQ? imaginary    ->    discrete 

\begin{figure}[tbh]%[htbp]
\begin{center}
\begin{tikzpicture}[decoration={
markings,mark=between positions 0.15 and 0.95 step 0.85cm with {\arrow{latex%stealth
}}}]
\def\R{2} % sphere radius
\def\S{5}%separation between shpere
\fill[fill=gray, opacity=0.1] (\R,0) circle (\R cm);
\draw[postaction={decorate},thick] (2*\R,0) arc  (0:180:\R cm);
\draw[postaction={decorate},thick] (0,0) arc  (180:360:\R cm);
\draw[postaction={decorate}] (1.8*\R,0) arc  (0:180:0.8*\R cm);
\draw[postaction={decorate}] (0.2*\R,0) arc  (180:360:0.8*\R cm);
\draw[postaction={decorate}] (1.6*\R,0) arc  (0:180:0.6*\R cm);
\draw[postaction={decorate}] (0.4*\R,0) arc  (180:360:0.6*\R cm);
\draw[postaction={decorate}] (1.4*\R,0) arc  (0:180:0.4*\R cm);
\draw[postaction={decorate}] (0.6*\R,0) arc  (180:360:0.4*\R cm);
\draw[postaction={decorate}] (1.2*\R,0) arc  (0:180:0.2*\R cm);
\draw[postaction={decorate}] (0.8*\R,0) arc  (180:360:0.2*\R cm);
  \fill[fill=black] (\R,0) circle (1.2pt);
 \draw[postaction={decorate},thick,xshift=\S cm] (2*\R,0) arc  (0:180:\R cm);
\draw[postaction={decorate},thick,xshift=\S cm] (0,0) arc  (180:360:\R cm);
\fill[fill=gray, opacity=0.1,xshift=\S cm] (\R,0) circle (\R cm);
\draw[postaction={decorate},xshift=\S cm] (\R,\R) arc  (90:270:0.9*\R cm);
\draw[postaction={decorate},xshift=\S cm] (\R,-0.8*\R) arc  (-90:90:0.9*\R cm);
\draw[postaction={decorate},xshift=\S cm] (\R,\R) arc  (90:270:0.8*\R cm);
\draw[postaction={decorate},xshift=\S cm] (\R,-0.6*\R) arc  (-90:90:0.8*\R cm);
\draw[postaction={decorate},xshift=\S cm] (\R,\R) arc  (90:270:0.65*\R cm);
\draw[postaction={decorate},xshift=\S cm] (\R,-0.3*\R) arc  (-90:90:0.65*\R cm);
\draw[postaction={decorate},xshift=\S cm] (\R,\R) arc  (90:270:0.5*\R cm);
\draw[postaction={decorate},xshift=\S cm] (\R,0) arc  (-90:90:0.5*\R cm);
\draw[postaction={decorate},xshift=\S cm] (\R,\R) arc  (90:270:0.3*\R cm);
\draw[postaction={decorate},xshift=\S cm] (\R,0.4*\R) arc  (-90:90:0.3*\R cm);
\draw[postaction={decorate},xshift=\S cm] (\R,\R) arc  (90:270:0.1*\R cm);
\draw[postaction={decorate},xshift=\S cm] (\R,0.8*\R) arc  (-90:90:0.1*\R cm);
  \fill[fill=black,xshift=\S cm] (\R,\R) circle (1.2pt);
\draw[postaction={decorate},thick,xshift=2*\S cm] (0,0) arc  (180:0:\R cm);
\draw[postaction={decorate},thick,xshift=2*\S cm] (0,0) arc  (180:360:\R cm);
\fill[fill=gray, opacity=0.1,xshift=2*\S cm] (\R,0) circle (\R cm);
\draw[postaction={decorate},xshift=2*\S cm] (0,0) to [out=85, in=180] (\R,0.85*\R) to [out=0, in=95](2*\R,0);
\draw[postaction={decorate},xshift=2*\S cm] (0,0) to [out=70, in=180] (\R,0.65*\R) to [out=0, in=110](2*\R,0);
\draw[postaction={decorate},xshift=2*\S cm] (0,0) to [out=45, in=180] (\R,0.45*\R) to [out=0, in=135](2*\R,0);
\draw[postaction={decorate},xshift=2*\S cm] (0,0) to [out=20, in=180] (\R,0.2*\R) to [out=0, in=160](2*\R,0);
\draw[postaction={decorate},xshift=2*\S cm] (0,0) to [out=0, in=180] (\R,0) to [out=0, in=180](2*\R,0);
\draw[postaction={decorate},xshift=2*\S cm] (0,0) to [out=-85, in=180] (\R,-0.85*\R) to [out=0, in=-95](2*\R,0);
\draw[postaction={decorate},xshift=2*\S cm] (0,0) to [out=-70, in=180] (\R,-0.65*\R) to [out=0, in=-110](2*\R,0);
\draw[postaction={decorate},xshift=2*\S cm] (0,0) to [out=-45, in=180] (\R,-0.45*\R) to [out=0, in=-135](2*\R,0);
\draw[postaction={decorate},xshift=2*\S cm] (0,0) to [out=-20, in=180] (\R,-0.2*\R) to [out=0, in=-160](2*\R,0);
   \fill[fill=black,xshift=2*\S cm] (0,0) circle (1.2pt);
      \fill[fill=black,xshift=2*\S cm] (2*\R,0) circle (1.2pt);
\node[below ] at (\R,-.15\textwidth) {$R: c^{(2)}=1$};  
\node[below ] at (\R+\S,-.15\textwidth) {$H: c^{(2)}=0$};  
\node[below ] at (\R+2*\S,-.15\textwidth) {${\bar R}: c^{(2)}=-1$};
\end{tikzpicture}
\caption{Time translation on the Poincar\'e disk. On the boundary of the disk (thick line), ``time flow'' is uniform without fixed point for $R$ or $c^{(2)}=1$ case, while it is limited to the finite region bounded by the two fixed points for  $\bar R$ or $c^{(2)}=-1$ case. $H$ or $c^{(2)}=0$ case exhibits marginal behavior, and it has one fixed point at infinity; The connection to dipolar quantization is apparent in this depiction.} \label{fig:poincaredisk}
\end{center}
\end{figure}
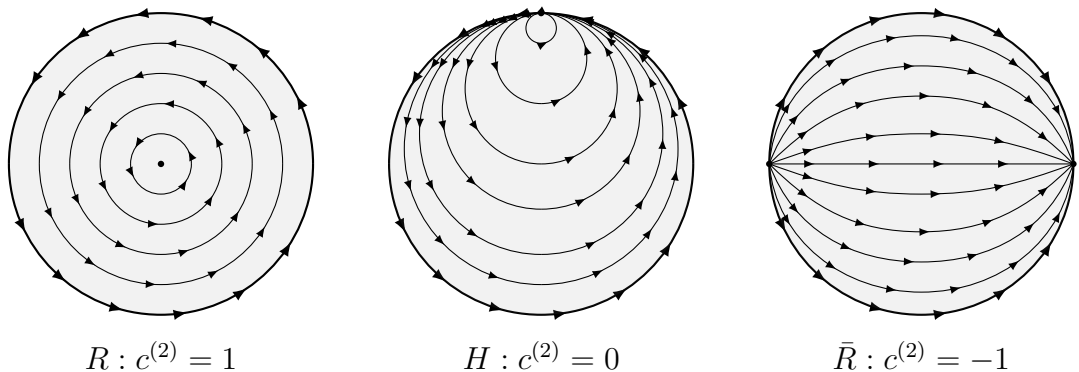

At this point, it would be insightful to contemplate the action of $sl(2, \mathbb{R})$. The $sl(2, \mathbb{R})$ algebra is also the Lie algebra of the projective special linear group $PSL(2,\mathbb R)=SL(2, \mathbb R)/\{\pm\}$ which is apparently a subgroup of  $SL(2, \mathbb R)$. The relation between $PSL(2,\mathbb R)$ and $SL(2, \mathbb R)$ is  reminiscent of the relation between $SO(3)$ and $SU(2)$. $PSL(2,\mathbb R)$ naturally acts on the hyperbolic plane $\mathbb H^{2}$, which is the upper half of the complex plane $\{z \in \mathbb C; \text{Im} z >0\}$ with the Poincar\'e metric
\begin{equation}
ds^{2}=\frac{|dz|^{2}}{(\text{Im} z)^{2}},
\end{equation}
or the Poincar\'e disk with the metric
\begin{equation}
ds^{2}=\frac{|dz|^{2}}{(1-|z|^{2})^{2}}.
\end{equation}
The action of $PSL(2,\mathbb R)$ on $\mathbb H^{2}$ gives the following automorphism:
\begin{equation}
z \mapsto \frac{az+b}{cz+d} ,  
\end{equation}
where $a,b,c,d \in \mathbb{R}$ and $ad-bc \neq 0$.
The action of $PSL(2,\mathbb R)$ on the Poincar\'e disk that corresponds to $R, H,$ and $\bar R$ respectively, is depicted in Fig. \ref{fig:poincaredisk}.
%\begin{figure}[tbh]
%\centering
%\begin{tikzpicture}
%\node[inner sep=0pt] (PD1) at (0,0)
%    {\includegraphics[width=.3\textwidth]{PD1.pdf}};
%\node[below ] at (0,-.15\textwidth) {$R: c^{(2)}=1$};
%\node[inner sep=0pt] (PD0) at (.33\textwidth,0)
%    {\includegraphics[width=.3\textwidth]{PD0.pdf}};
%\node[below ] at (.33\textwidth,-.15\textwidth) {$H: c^{(2)}=0$};
%\node[inner sep=0pt] (PD-1) at (.66\textwidth,0)
%    {\includegraphics[width=.3\textwidth]{PD-1.pdf}};
%\node[below ] at (.66\textwidth,-.15\textwidth) {${\bar R}: c^{(2)}=-1$};
%\end{tikzpicture}
%\caption{Time translation on the Poincar\'e disk. On the boundary of the disk, ``time flow'' is uniform without fixed point for $R$ or $c^{(2)}=1$ case, while it is limited to the finite region bounded by the two fixed points for  $\bar R$ or $c^{(2)}=-1$ case. $H$ or $c^{(2)}=0$ case exhibits marginal behavior, and it has one fixed point at infinity; The connection to dipolar quantization is apparent in this depiction.} \label{fig:poincaredisk}
%\end{figure}
Note, in particular, that  the M\"obius transformation is similar to the transformation above, except that it forms a complex Lie group and is isomorphic to the automorphism of the Riemann sphere $\text{Aut}(  \widehat{\mathbb C})$ rather than to the automorphism of the half plane. 
%$\mathbb H^{2}$ $\text{Aut}( {\mathbb H}^{2})$. 
%Added April 10
A distinct feature that one can discern from Fig. \ref{fig:poincaredisk} is the flow on the edge of the disk, which consists the one-dimensional coordinate ``t''. The time for $R$ and  $H$ respectively, flows in one direction and encircles the entire edge. On the other hand, the flow for $\bar R$ is divided into the two segments on the edge. There, the time flow in one segment covers only half of the entire edge. The situation parallels the case studied by Bisognano and Wichmann \cite{Bisognano:1975ih}, where the modular Hamiltonian (automorphism, to be exact) maps the half of the entire space into itself. This resemblance further supports the interpretation presented above that $\bar R$ corresponds to the modular Hamiltonian, or the entanglement Hamiltonian. Also, one may consider the other half "space" which cannot be reached by the time flow as the space effectively integrated out. This consideration further supports the above mentioned connection to the entanglement Hamiltonian.
The observation here would  also be useful in the study of  SSD for the case of open strings, where the setup of the upper half plane is natural.

In summary, we find the same structure in CQM as observed in 2d CFT where the choice of the Hamiltonian leads to radial quantization, the dipolar quantization or SSD, and the entanglement Hamiltonian, respectively. We identify the respective Hamiltonians in CQM using $sl(2, \mathbb{R})$ symmetry. The findings here will offer a simpler setup for the study of SSD and the entanglement Hamiltonians.
It would be also interesting to investigate further  in the context of the conformal boot strap approach \cite{Qiao:2017xif} or the recent discussion of  the CQM correlation function \cite{Khodaee:2017tbk}.

\vspace{0.2cm}
\noindent{\bf Acknowledgement:} The author would like to thank N. Ishibashi, H. Katsura, H. Kawai, K. Okunishi, S. Ryu, and the participants of the iTHEMS workshop "Workshop on Sine square deformation and related topics, " for fruitful discussions and many suggestions, which greatly contributed to the present work.

%\begin{figure}[tbh]
%\centering
%\includegraphics[width=15cm]{  .pdf}
%\caption{
%} \label{fig: }
%\end{figure}

\end{document}